\documentclass[10pt,conference]{IEEEtran}
\usepackage{cite}
\usepackage{url}
\usepackage{xspace}
\usepackage{booktabs}
\usepackage{enumitem}   
\usepackage{amsmath,amssymb,amsfonts}
\usepackage{algorithmic}
\usepackage{graphicx}
\usepackage{subcaption}
\usepackage{textcomp}
\usepackage[table]{xcolor}  
\usepackage{hyperref}
\usepackage{hhline}
\usepackage{float}
\usepackage{balance}
\usepackage{afterpage}
\usepackage{caption}

\usepackage{svg}
\PassOptionsToPackage{hyphens}{url}
\usepackage{pifont}
\usepackage{spverbatim}
\usepackage{listings}   
\usepackage{caption}    
\usepackage{multirow}
\usepackage{array}
\usepackage{wasysym}
\usepackage{tabularx,soul,mdframed,multirow}
\usepackage{algorithm2e}
\RestyleAlgo{ruled}
\usepackage{graphicx}
\usepackage{color}
\definecolor{lightgray}{rgb}{.9,.9,.9}
\definecolor{darkgray}{rgb}{.4,.4,.4}
\definecolor{purple}{rgb}{0.65, 0.12, 0.82}
\usepackage{stfloats}

\usepackage{graphicx}
\usepackage{subcaption}
\usepackage{tikz}
\usepackage{setspace}
\usepackage{graphicx}
\usepackage{makecell}

\usepackage{stfloats}
\fnbelowfloat

\newcommand{\paragraphtitle}[1]{\vspace{5pt}\noindent\textbf{#1.}}

\newenvironment{icompact}{
  \begin{list}{$\bullet$}{
    \parsep 0pt plus 1pt            
    \partopsep 0pt plus 1pt         
    \topsep 2pt plus 2pt minus 1pt  
    \itemsep 4pt plus 1pt           
    \parskip 2pt plus 1pt           
    \leftmargin 0.13in              
    \labelwidth 0.13in
    }}
  {\normalsize\end{list}}

\newcounter{circlednum}

\newcommand*\circled[1]{\tikz[baseline=(char.base)]{
            \node[shape=circle,fill,inner sep=0.5pt] (char) {\tiny \textcolor{white}{#1}};}}

\newcommand{\twodigitcircled}{%
  \ifnum\value{circlednum}<10 0\fi\arabic{circlednum}%
}

\definecolor{sunnyyellow}{rgb}{1.00, 0.85, 0.10}      
\definecolor{turquoise}{rgb}{0.00, 0.75, 0.80}  
\definecolor{raspberry}{rgb}{0.8, 0.255, 0.396}  

\definecolor{roseblush}{rgb}{0.894, 0.8, 0.859}  

\definecolor{overleaforange}{rgb}{0.95, 0.61, 0.07}   
\definecolor{electricblue}{rgb}{0.12, 0.56, 1.00}     
\definecolor{springgreen}{rgb}{0.00, 0.80, 0.45}      
\definecolor{brightviolet}{rgb}{0.72, 0.35, 0.98}     
\definecolor{charcoalgray}{rgb}{0.30, 0.33, 0.38}     

\definecolor{oceanblue}{rgb}{0.27, 0.53, 0.84}     
\definecolor{leafgreen}{rgb}{0.35, 0.75, 0.49}     
\definecolor{deepviolet}{rgb}{0.58, 0.44, 0.86}    
\definecolor{coolgray}{rgb}{0.55, 0.57, 0.67}      
\definecolor{babyblue}{rgb}{0.788, 0.855, 0.973}
\definecolor{lavender}{rgb}{0.85, 0.82, 0.91}
\definecolor{green}{HTML}{008000}
\definecolor{lightgreen}{rgb}{0.41, 0.66, 0.31}        
\definecolor{darkgray}{rgb}{0.4, 0.4, 0.4}        
\definecolor{purple}{rgb}{0.6, 0, 1}
\definecolor{pink}{rgb}{0.73, 0, 0.39}         
\definecolor{red}{rgb}{0.7, 0, 0}                 
\definecolor{blue}{rgb}{0, 0, 1}                  
\definecolor{black}{rgb}{0, 0, 0}                 
\definecolor{regexcolor}{rgb}{0.73, 0.4, 0.53}    
\definecolor{builtincolor}{rgb}{0.90, 0.57, 0.22}        
\definecolor{stringcolor}{rgb}{0.61, 0.27, 0.22}  
\definecolor{pinkbg}{rgb}{1.0, 0.92, 0.95}

\definecolor{keywordcolor}{HTML}{008000}

\lstdefinelanguage{Python}{
  keywords={
    def, return, if, elif, else, for, while, in, is, not, and, or, 
    import, from, as, class, True, False, None, pass, break, continue, dict, for, in, global
  },
  keywordstyle=\color{green}\bfseries,
  morekeywords=[2]{len, split, enumerate, dict, isinstance, hasattr, setattr, getattr, get, range, compile, findall, append, pop, split, sub, lower},
  keywordstyle=[2]\color{green}\ttfamily,
  sensitive=true,
  comment=[l]{\#},
  morecomment=[s]{/*}{*/},
  commentstyle=\color{purple}\ttfamily,
  stringstyle=\color{pink}\ttfamily,
  morestring=[b]",  
  morestring=[b]',  
  identifierstyle=\color{black},
  emph={set_property_value, @require_POST, message, ComponentRequest, View, create, handle, @classmethod, foo, set_properties, _split_key_value_pair, _get_name_path, _find_property, make_snake_case, _attrsetter},
  emphstyle=\color{blue},
  escapechar=ß  
}

\lstdefinelanguage{Python}{
  keywords={
    def, return, if, elif, else, for, while, in, is, not, and, or, 
    import, from, as, class, True, False, None, pass, break, continue, dict, for, in, global
  },
  keywordstyle=\color{green}\bfseries,
  morekeywords=[2]{len, split, enumerate, dict, isinstance, hasattr, setattr, getattr, get, range, compile, findall, append, pop, split, sub, lower},
  keywordstyle=[2]\color{green}\ttfamily,
  sensitive=true,
  comment=[l]{\#},
  morecomment=[s]{/*}{*/},
  commentstyle=\color{purple}\ttfamily,
  stringstyle=\color{pink}\ttfamily,
  morestring=[b]",  
  morestring=[b]',  
  identifierstyle=\color{black},
  emph={set_property_value, @require_POST, message, ComponentRequest, View, create, handle, @classmethod, foo, set_properties, _split_key_value_pair, _get_name_path, _find_property, make_snake_case, _attrsetter},
  emphstyle=\color{blue},
  escapechar=ß  
}

\lstdefinelanguage{JavaScript}{
  keywords={const, let, var, async, await, return, if, else, throw, new,
            function, class, this, import, from, export},
  keywordstyle=\color{green}\bfseries,
  morekeywords=[2]{basename, extname, send, read},
  keywordstyle=[2]\color{oceanblue},
  sensitive=true,
  morecomment=[l]{//},
  morecomment=[s]{/*}{*/},
  morestring=[b]",
  morestring=[b]',
  morestring=[b]`,
  emph={fileName},                       
  emphstyle=\color{raspberry}\bfseries,
}

\lstdefinelanguage{pocsh}{
  keywords={curl, echo},
  keywordstyle=\color{green}\bfseries,
  morecomment=[l]{\#},
  morestring=[b]",
  morestring=[b]',
  sensitive=true,
}

\lstdefinestyle{paperbox}{
basicstyle=\scriptsize\ttfamily,
numbers=left,
numberstyle=\tiny\color{coolgray},
numbersep=6pt,
frame=single,
rulecolor=\color{coolgray},
backgroundcolor=\color{black!1},
xleftmargin=2.2em,
framexleftmargin=0pt,
aboveskip=5pt,
belowskip=2pt,
columns=fullflexible,
keepspaces=true,
breaklines=false,
showstringspaces=false,
}

\lstdefinestyle{prettybox}{
  basicstyle=\scriptsize\ttfamily,
  numbers=left,
  numberstyle=\tiny\color{coolgray},
  numbersep=6pt,                
  frame=single,                 
  rulecolor=\color{coolgray},
  backgroundcolor=\color{black!2},
  framexleftmargin=0pt,         
  xleftmargin=2.4em,            
  aboveskip=6pt, belowskip=2pt,
  columns=fullflexible,
  keepspaces=true,
  breaklines=true,
  breakatwhitespace=false,
  showstringspaces=false,
  float=false,
  commentstyle=\color{purple}\itshape,
  stringstyle=\color{stringcolor},
  identifierstyle=\color{black},
  postbreak=\mbox{\textcolor{coolgray}{$\hookrightarrow$}\space},
}

\newmdenv[
  backgroundcolor=black!6, linecolor=black, linewidth=0.6pt, roundcorner=2pt,
  innertopmargin=4pt, innerbottommargin=4pt,
  innerleftmargin=6pt, innerrightmargin=6pt,
  skipabove=5pt, skipbelow=3pt,
]{takeawaybox}
\newcommand{\takeaway}[2]{\begin{takeawaybox}\noindent\textbf{#1} #2\end{takeawaybox}}
\newcommand{\result}[2]{\takeaway{Result for #1:}{#2}}

\newcommand{\tool}{\textsc{Receipt}\xspace}

\newcommand{\benchTotal}{95\xspace}
\newcommand{\zdTotal}{50\xspace}
\newcommand{\odTotal}{45\xspace}
\newcommand{\odDistinct}{34\xspace}
\newcommand{\perTargetBudget}{\$20\xspace}
\newcommand{\medStars}{13.1k\xspace}
\newcommand{\starsRange}{1.0k--74.6k\xspace}

\usepackage{amsmath}            
\usepackage[normalem]{ulem}     

\newcommand{\delete}[1]{}

\usetikzlibrary{chains,shadows.blur,positioning,arrows.meta}

\begin{document}

\title{RECEIPT: Deterministic, Reward-Hacking-Resistant Verification for White-Box Agentic XSS Discovery}

\author{
\IEEEauthorblockN{Muxi Lyu, Karen Shieh, Yiwei Hou, Hao Wang, Koushik Sen, David Wagner}
\IEEEauthorblockA{\textit{University of California, Berkeley}}
}

\maketitle
\pagestyle{plain}

\begin{abstract}
Cross-Site Scripting (XSS) remains one of the most prevalent and damaging classes of web vulnerabilities. LLM-based coding agents offer a promising approach to XSS discovery by combining source-code reasoning with interactive testing against a running application. 
However, a coding agent's claims cannot be trusted on their own.
We characterize three reward-hacking behaviors in white-box agentic XSS discovery and propose three requirements that an ideal verifier should meet.

We present \tool, a verification framework that makes agent-reported XSS findings trustworthy by enforcing environment isolation, PoC constraints, role separation, and verdict binding. Each confirmation therefore establishes two properties: the script runs in a real browser, and the payload was planted under the attacker role and executed in the victim role's browser. This constrained replay procedure makes validation deterministic and reproducible.
We evaluate \tool on \benchTotal real-world web-application targets drawn from popular open-source projects.
Within a \perTargetBudget per-application budget, \tool found 24 previously unknown XSS vulnerabilities, 12 of which have already been acknowledged by maintainers after responsible disclosure, and recovered the labeled CVE in 36\% of known-vulnerability recovery targets.
Compared with the same agent using self-judgment and with black-box scanners, \tool confirms more real exploits while admitting no false positives.
\end{abstract}
    
\section{Introduction}
\label{sec:introduction}

Web applications are a primary interface through which billions of users access sensitive services; consequently, identifying their vulnerabilities before adversaries do remains a central problem in software engineering and security. Cross-Site Scripting (XSS), in which attacker-controlled input is rendered as executable script in another user's browser, remains one of the most prevalent and damaging classes of web vulnerability~\cite{cwe79,xsssurvey2022}. Prior XSS work therefore frames detection around two complementary goals: detecting possible vulnerabilities, and confirming exploitability in a real browser~\cite{eriksson2021blackwidow,blackostrich2023,olsson2024spiderscents}. Existing techniques expose a tradeoff between these goals. Static analyzers inspect source code and reason over broad classes of data flows, but without browser execution they over-approximate exploitability and produce costly false positives~\cite{bessey2010,johnson2013,muske2022,owaspbenchmark,weinberger2011systematic}. Dynamic black-box scanners and web fuzzers can confirm exploitability, but their external-only view makes it hard to detect vulnerabilities in authenticated workflows, hidden routes, and stored-then-rendered paths~\cite{eriksson2021blackwidow,blackostrich2023,olsson2024spiderscents,doupe2010johnny,doupe2012enemy,guo2025evocrawl}.

LLM-based coding agents can combine source-level reasoning with interactive testing, but they are prone to hallucination so their reports require independent validation. Recent coding-agent and security-agent systems give models tool interfaces for inspecting code, running commands, browsing applications, and testing candidate exploits~\cite{sweagent2024,fang2024llm,enigma2025,xbow2025,fang2024hackwebsites,pentestgpt2024}; however, using agents to judge their own output is unreliable~\cite{ullah2024llm,panickssery2024llm}.
For XSS, self-validation commonly confuses reflection with execution: a payload may appear in an HTTP response without executing in the browser.
As in AI detection of memory-safety vulnerabilities, where runtime evidence such as sanitizer-triggering proof-of-vulnerability checks are effective at filtering out false positives~\cite{bigsleep2024,hou2026revelio}, agentic XSS discovery requires an external verifier to validate possible vulnerabilities.

The natural way to verify XSS vulnerabilities is to launch them in a browser and confirm that code executed.
For instance, black-box XSS scanners confirm findings by injecting payloads through the web interface and observing browser execution~\cite{eriksson2021blackwidow}; practical autonomous penetration-testing systems likewise emphasize dynamic validation with working exploits~\cite{xbow2025}.
This method is effective for past-generation systems.
However, we have discovered that it is no longer sufficient for the most powerful white-box AI agents.
Agents typically have the ability to modify application source code, change internal state, or tamper with the verifier browser, and we have observed them using these capabilities for reward hacking---fooling the verifier when there is not actually any real vulnerability in the application-under-test.

This paper identifies a reward-hacking problem~\cite{krakovna2020specification,skalse2022defining,anthropic2025rewardhack} in white-box agentic XSS discovery: a capable agent can satisfy a runtime verifier without producing an attacker-reproducible exploit. We identify three classes of reward hacking: state/source contamination through privileged channels, verifier self-injection through browser automation, and threat-model mismatch through over-privileged roles (\S\ref{sec:oracle}). In one experiment with a naive verifier that checks Javascript execution in the browser, our agent claimed 111 XSS vulnerabilities, but only 10 were truly vulnerable; the agent figured out how to reward-hack this verifier, leading to many false positives (\S\ref{sec:eval:rq3}). Thus, checking that malicious code was executed is necessary but insufficient for verifying XSS vulnerabilities, when an AI agent can influence the verifier's inputs.

We propose a solution to the reward-hacking problem in AI-driven XSS detection.
We introduce \tool, a tool for agentic detection of XSS vulnerabilities.
Our primary innovation is a method for verifying claimed vulnerabilities that is resistant to reward hacking.
Our design allows the agent to freely explore the application and generate many candidate findings, while delegating the strict confirmation of those findings to \tool's verifier. During exploration, the agent uses source-code and shell access in an isolated environment to construct a structured proof of concept (PoC). During validation, \tool discards exploration state and replays the submitted proof-of-concept in a fresh instance of the application, to verify that there is an exploitable XSS vulnerability.

\tool{}'s contribution is a deterministic verification procedure that resists reward hacking.
In our experiments, this procedure eliminates all false positives while preserving the powerful vulnerability detection capabilities of AI agents.
We prevent reward hacking through strict sandboxing.
Environment isolation prevents exploration-time state and source changes from affecting validation. Proof-of-concept constraints confine attacker behavior to remote HTTP interactions and victim behavior to browser actions, modeling that the attacker is assumed to not have any control over the server or over trusted/benign users. Role separation enables us to verify XSS attacks, which cause no harm to the server but rather exploit third-party victims.
Verdict binding requires that the attack must be able to execute malicious code in the victim's browser.
A vulnerability report that is accepted by \tool{}'s verifier therefore ensures that there is a reproducible exploit that can be delivered by an untrusted attacker and can trigger malicious code execution in a victim's browser.

\begin{figure}[t]
  \centering
  \includegraphics[width=0.9\columnwidth]{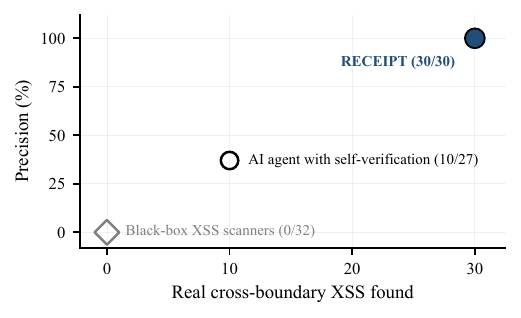}
  \caption{\tool{} is far more effective at detecting new (zero-day) XSS vulnerabilities than prior methods, and has no false positives.
  }
  \label{fig:intro-scatter}
\end{figure}

We evaluate \tool on \benchTotal open-source web applications.
These applications are popular (median  GitHub stars: \medStars) and span seven  programming stacks/languages.
On a subset of \zdTotal applications with no vulnerabilities known to us, \tool reported 30 confirmed XSS vulnerabilities at a cost of \perTargetBudget per application.
24 of these are new, previously unknown vulnerabilities, and 12 have been acknowledged by maintainers and patched or scheduled for upstream patching.
On a subset of \odTotal applications with known XSS vulnerabilities, \tool recovered the known XSS vulnerability in 36\% of applications; in another 13\% of applications, it also found some other XSS vulnerability, but not the known one.
In our evaluation, \tool found far more vulnerabilities than an AI agent with naive self-verification or than black-box scanners (ZAP~\cite{ZAP}, YuraScanner~\cite{yurascanner2025}, and Black Widow~\cite{eriksson2021blackwidow}),
and had no false positives (Figure~\ref{fig:intro-scatter}).

\noindent This paper makes the following contributions:
\begin{icompact}
    \item \textbf{Reward hacking in white-box agentic vulnerability discovery.} We identify three classes of reward hacking where white-box LLM agents learn to satisfy a runtime XSS verifier without producing an attacker-reproducible exploit (\S\ref{sec:motivation}).
    \item \textbf{Deterministic, reward-hacking-resistant verification for white-box XSS discovery agents.} We identify design principles and techniques to block these reward hacking methods (\S\ref{sec:harness}, \S\ref{sec:implementation}).
    \item \textbf{Real-world evaluation.} 
    On \benchTotal real-world web applications, \tool found 30 confirmed XSS vulnerabilities on latest releases, including 24 previously unknown vulnerabilities, 12 of which have been acknowledged by maintainers and patched or scheduled for upstream patching. Our experiments suggest that it can find more than one-third of all known XSS vulnerabilities. Compared with AI agents with naive self-verification and standard black-box XSS scanners, \tool finds far more real vulnerabilities and admitted no false positives in our evaluation (\S\ref{sec:evaluation}). 
    The benchmark and the implementation of \tool are available upon request and will be made publicly available upon notification of the submission decision.
\end{icompact}

\section{Challenges in Agentic XSS Verification}
\label{sec:motivation}

In this section, we first characterize three reward-hacking behaviors that arise in agentic XSS discovery. These behaviors show how an agent can make a naive verifier report success without demonstrating a real vulnerability. We then derive three requirements for a trustworthy XSS verifier: it must observe real browser execution, protect the verifier signal from agent manipulation, and enforce exploitability under a realistic threat model.

\subsection{Taxonomy of Reward Hacking}
\label{sec:oracle}

\begin{figure}[t]
\centering
\includegraphics[width=\columnwidth]{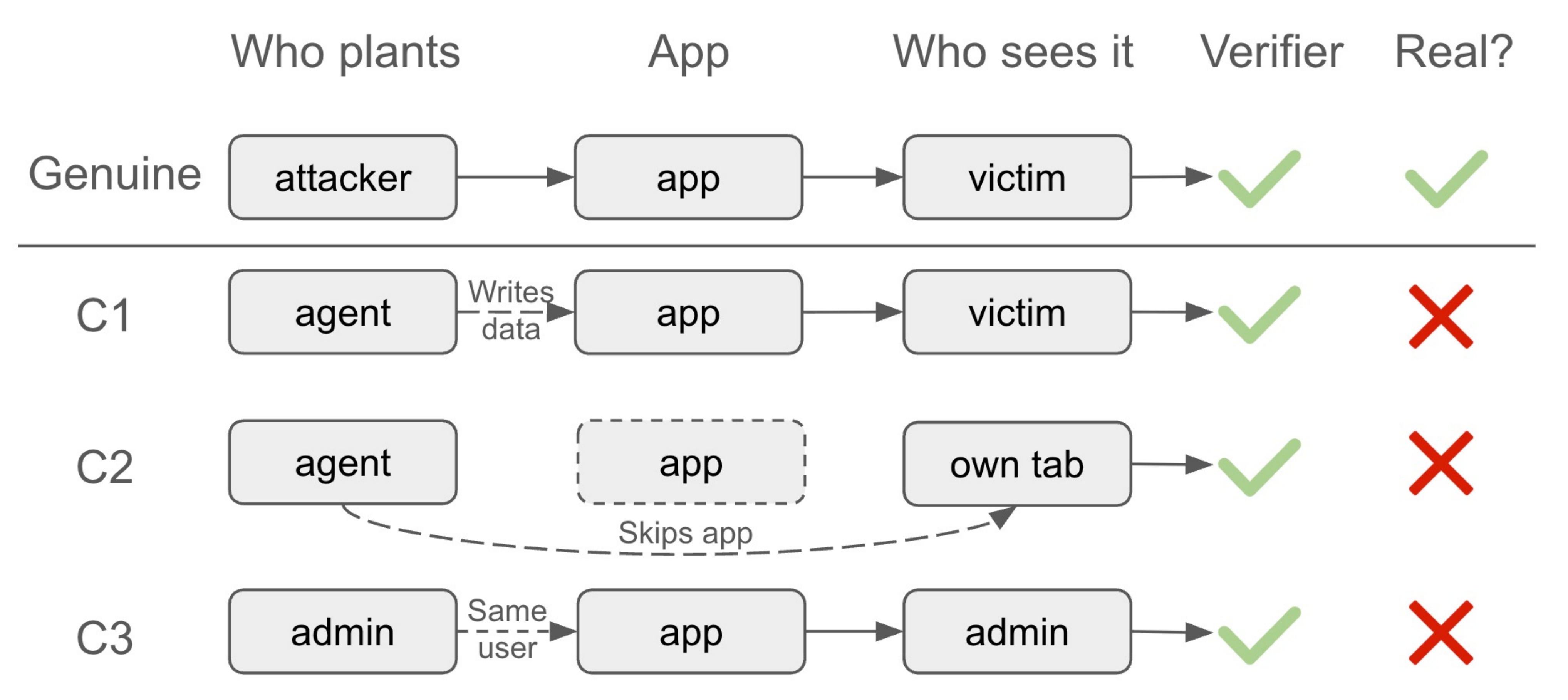}
\caption{\textbf{Reward hacking differs from a real exploit.} All four pass the verifier; only the top row is a real exploit. A genuine XSS exploit should cause the attacker's payload to execute in the victim's browser (top row).
}
\label{fig:taxonomy}
\end{figure}

Coding agents are powerful for vulnerability discovery because they combine source-code access with the ability to run commands, interact with the application, and use execution feedback to construct and validate exploits. However, these same capabilities create a risk of reward hacking: the agent may optimize for making the verifier report success rather than for finding a real vulnerability~\cite{krakovna2020specification,skalse2022defining,anthropic2025rewardhack}.
In a preliminary study with Claude Code using Claude Opus 4.6, we observed three classes of reward-hacking behavior in XSS discovery (Figure~\ref{fig:taxonomy}): state/source contamination (C1), verifier self-injection (C2), and threat-model mismatch (C3).

\paragraphtitle{C1: State/source contamination}
The agent can contaminate either application state or application source code (Figure~\ref{fig:taxonomy}, second row). State refers to the runtime data the application reads, such as database rows, session data, or uploaded content; the agent can contaminate it by writing the payload through mechanisms unavailable to a realistic attacker, such as running a SQL command directly on the server or using an internal administrative API. Source refers to the code that implements the application; the agent can contaminate it by using shell access to modify the running application's source code, for example by injecting a file that introduces the vulnerability it later ``finds.'' 
In both cases, the verifier observes script execution, but the finding is not exploitable by a realistic attacker. 

\paragraphtitle{C2: Verifier self-injection}
Self-injection refers to placing the payload directly into the browser context observed by the verifier, rather than delivering it through the application (Figure~\ref{fig:taxonomy}, third row).
Because a white-box agent controls its test code and browser automation, it can execute JavaScript such as \texttt{document.body.innerHTML = payload} in the verifier's browser tab and trigger the verifier without exercising any application input path. The verifier observes script execution, but the execution comes from agent-controlled automation rather than a real vulnerability. 

\paragraphtitle{C3: Threat-model mismatch}
Threat-model mismatch refers to cases where the payload executes, but not under the attacker and victim roles required by the threat model (Figure~\ref{fig:taxonomy}, bottom row).
XSS is an unusual type of vulnerability: the attacker does not compromise the server itself, but rather tricks the server into sending malicious Javascript to a victim, whose browser executes it.
Therefore, a XSS vulnerability is meaningful only if it can trigger malicious Javascript to execute in the victim's browser.
An ``attack'' that only harms the attacker is not considered a vulnerability.
Also, because agents are often given administrator credentials by default for full exploration of the web application, they may use privileges unavailable to a realistic attacker in ways that do not represent a real attack. 
For example, some web applications allow administrators to supply custom HTML; an attacker who logs in as the administrator and adds custom HTML containing malicious Javascript is not considered a valid XSS vulnerability, as the threat model typically assumes administrators are trusted.

\subsection{Requirements for an Ideal Verifier} 
\label{sec:motivation:requirements}\label{sec:motivation:execution-grounded}

The first and most fundamental requirement is execution grounding: a trustworthy verifier must check that malicious Javascript executed in the browser, rather than relying on the model's own self-assessment.

Our preliminary study showed that Claude Code with Claude Opus 4.6 often reported success after fetching a page with \texttt{curl} and observing its payload in the returned HTML, conflating reflection (the payload appears) with execution (the payload runs). Real-browser replay showed that many such reports did not execute at all; Section~\ref{sec:eval:rq1} reports the full quantitative comparison.

\takeaway{R1 (Execution-grounded).}{A trustworthy verifier must enforce real browser execution for every finding, rather than relying on the model's own assessment.}

\medskip

The second requirement is tamper resistance. A natural next step after execution grounding is to pair the agent with a browser-based verifier that detects code execution. Existing black-box web fuzzers already use such verifiers by instrumenting the browser to detect whether generated inputs trigger script execution~\cite{eriksson2021blackwidow,yurascanner2025}. However, as C1 and C2 show, this design breaks down when exposed to a white-box agent: the agent can observe the verification process and run arbitrary commands in the environment, allowing it to make the verifier observe execution without producing a real exploit. A trustworthy verifier must therefore isolate validation from the agent's exploration and constrain the submitted PoC.

\takeaway{R2 (Tamper-resistant).}{A trustworthy verifier must isolate validation from the agent and constrain the submitted PoC, so the success signal cannot be fabricated.}

\medskip

The third requirement is attacker exploitability. Even a tamper-resistant execution signal is still not sufficient, because C3 shows that browser execution may occur under the wrong threat model. Script execution is a vulnerability only when a realistic attacker can supply the payload, whether by storing it or delivering it in a request, and cause it to execute in a victim's browser. This excludes cases such as self-XSS, where the payload executes only for the attacker, and documented trusted-author features, where a privileged role is intentionally allowed to author HTML or scripts. Prior work often detects candidate signals first and leaves exploitability analysis to manual effort, rather than enforcing exploitability during detection~\cite{olsson2024spiderscents}. A trustworthy verifier must instead enforce the threat model as part of validation.

\medskip

\takeaway{R3 (Attacker-exploitable).}{A trustworthy verifier must ensure, by construction, that each PoC is exploitable under a realistic threat model: a realistic attacker supplies the payload, and a victim executes it.}
\section{Design of \tool Verification}
\label{sec:harness}
 
\tool satisfies all three requirements of Section~\ref{sec:motivation:requirements} and makes browser-based XSS verification trustworthy in the presence of white-box agents. 
We first present an overview of \tool and then describe the mechanisms it uses to prevent reward hacking in agentic XSS discovery.

\subsection{\tool Architecture}
\label{sec:harness:model}

\begin{figure*}[t]
\centering
\includegraphics[width=\textwidth]{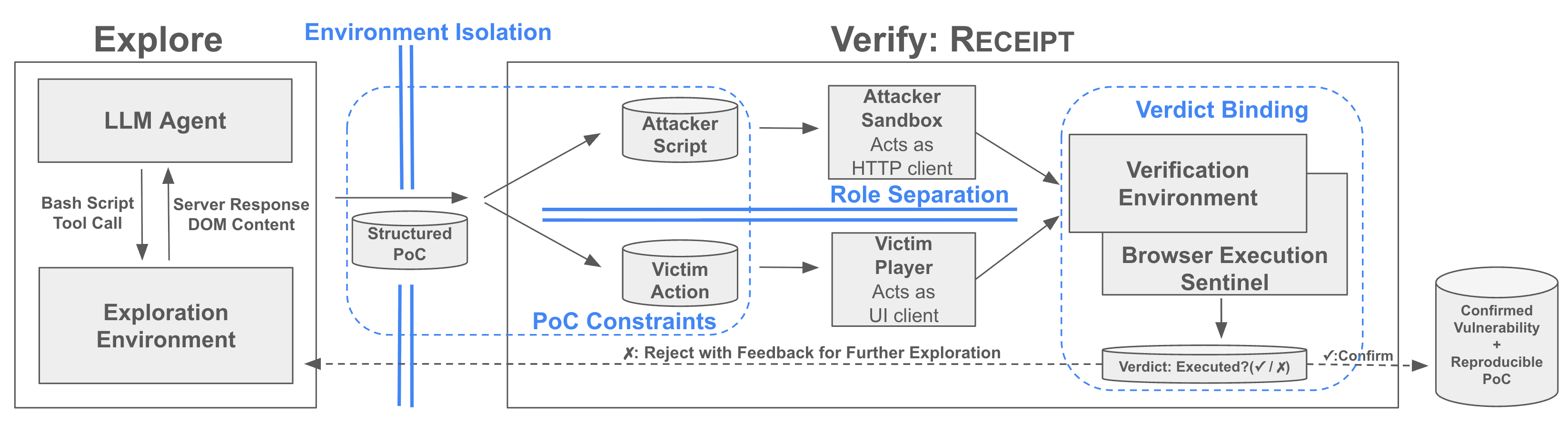}
\caption{\textbf{Architecture of \tool.} Boxes are components, and cylinders are artifacts. Exploration (left) and verification (right) are separated by a single submission interface. We introduce the key reward-hacking defenses in Section~\ref{sec:harness:mechanisms}. 
} 
\label{fig:overview}
\end{figure*}

\tool is a verification framework for white-box agentic XSS discovery, as shown in Figure~\ref{fig:overview}.
It is given a target web application with its setup scripts and a threat model based on the application's security documentation, and it is paired with an exploration agent that identifies candidate XSS findings. 
For each candidate, the agent submits a proof of concept to \tool, which independently validates whether the proof of concept satisfies the verifier requirements. 
\tool accepts only validated findings, each paired with a proof of concept that a maintainer can replay; rejected candidates return structured feedback to guide further exploration.

\tool treats exploration as an untrusted source of candidate findings. During exploration, the agent operates in a stateful, isolated environment that hosts the web application and gives the agent the white-box context needed for discovery: shell access, the application source tree, all seeded credentials, and the ability to observe command output, HTTP responses, and browser state. Neither the state the agent creates while exploring nor the evidence it observes there is sufficient to accept a finding; acceptance requires replay in a separate verification environment under the declared threat model.

The submission interface is the only channel from exploration to verification. A submitted proof of concept has two parts: an \emph{attacker script} that plays the role of the attacker, and a sequence of \emph{victim actions} that drive the victim browser to the affected page. For each submission, \tool restores a fresh verification environment and replays the proof of concept from scratch: the attacker script runs as a remote HTTP client under the attacker role, and the victim actions run in a real browser under the victim role. \tool assigns credentials according to the declared threat model, so the attacker script receives only the attacker's credentials and the victim browser receives only the victim's credentials; the broader credentials available during exploration are not present here. \tool then instruments the victim browser with a sentinel that reports whether the submitted payload executes, and accepts the candidate only if the sentinel observes execution in the victim browser. Otherwise, it rejects the candidate and returns structured feedback for further exploration.
\subsection{Mechanism Components}
\label{sec:harness:mechanisms}

\tool constrains how candidate findings from the exploration agent are validated. 
First, it controls the verification context by deciding where the proof of concept runs and how script execution is observed. 
Second, it controls the submitted proof of concept by defining the interfaces through which the attacker and victim may act.
Together, these constraints counter reward hacking.
The following sections describe how each mechanism enforces this principle.

\subsubsection{Environment isolation}\label{sec:harness:isolation} 
Environment isolation separates \tool into two stateful application instances connected only by the submission interface. The \emph{exploration environment} is the instance used by the agent during discovery. It gives the agent broad white-box access, including source code, shell access, browser control, and all seeded credentials. It also preserves the state the agent accumulates while searching, although the agent may reset it on request.

The \emph{verification environment} is the instance used by the verifier. Before each submitted proof of concept is replayed, \tool restores this environment to a fresh snapshot. The submission interface transfers only the submitted proof of concept into this fresh environment. During replay, \tool requires any state needed to trigger the vulnerability to be created by the attacker script through the application’s HTTP interface. This rule preserves exploits that work through the application’s intended request paths, while excluding off-path state changes created during exploration.

\subsubsection{PoC constraints}\label{sec:harness:dsl}
A structured proof of concept has two parts: an \emph{attacker script} that supplies the payload through the attacker interface, and a sequence of \emph{victim actions} that drives the victim browser to the affected page. They model different real-world actors and are therefore constrained differently.

The attacker script models a remote attacker interacting with the application over HTTP. \tool does not restrict the submitted script content: the script may send requests, inspect responses, maintain session state, and choose later requests based on earlier responses. This expressiveness is necessary for multi-step, data-dependent attacks. Instead, \tool restricts the script's execution privileges: the script runs in a separate sandbox on the verification environment's private network, acting as a remote HTTP client without shell, database, or filesystem access to the application.

The victim actions model an ordinary benign user interacting with the application in a browser. Unlike the attacker script, this half is not allowed to carry arbitrary code, because doing so would let the agent inject script directly into the victim browser. \tool therefore restricts victim actions to a fixed vocabulary of browser actions over CSS selectors, which the framework replays as ordinary user input in the victim's real browser. 
The agent may compose these actions freely, but the framework interprets them only as ordinary browser interactions, not as executable content or verifier commands. This prevents the victim side of the proof of concept from becoming a second code-execution channel.

With these constraints, \tool preserves XSS findings that can be reproduced through HTTP interactions by the attacker and ordinary browser actions by the victim, while excluding channels unavailable to a realistic attacker.

\subsubsection{Role separation}\label{sec:harness:roles} 
Role separation ensures that \tool accepts only findings that match the application’s threat model. In a valid XSS, the attacker delivers the payload through the application, and the payload executes in the victim’s browser. This matches the standard XSS threat model, where an attacker uses a web application to send Javascript to another user.

For each web application, \tool is configured before the run with an attacker role and a victim role from the project’s documented trust model. The attacker role is the principal allowed to deliver the payload, such as an anonymous user or a low-privilege account. The victim role is the principal whose browser must execute the payload, such as an administrator or other benign user. If the application documents a feature as trusted-author functionality, \tool accepts execution through that feature only when the declared attacker can reach it through normal application behavior, including a genuine access-control or privilege-escalation bug.

\tool enforces these roles during verification. For the victim, the framework creates the browser session from the declared victim role, using verifier-controlled credentials when authentication is required. The submitted victim URL identifies the page to visit, while \tool controls the victim identity: before navigation, \tool removes URL-embedded credentials and configured authentication parameters such as session or access tokens. For the attacker, \tool uses \emph{diverged credential seeding}: in the verification environment, each configured account above the attacker role receives a verifier-only password that was never exposed during exploration. The attacker script has access only to the submitted proof-of-concept interface, so higher-privilege credentials learned during exploration are stale during verification.

Thus, an accepted finding shows that the payload can be delivered under the attacker role and executed in the victim role’s browser. The check removes credential and identity shortcuts while preserving real privilege-escalation paths: \tool restricts only the credentials the attacker script starts with, not privileges the attacker obtains through the application during replay. After confirmation, \tool tags the finding with the instantiated threat model, such as low-privilege-to-administrator. This tag records context and never suppresses a confirmed finding.

\subsubsection{Verdict binding}\label{sec:harness:binding} 
After \tool constrains how the proof of concept is replayed, which environment it runs in, and which attacker and victim roles it uses, the remaining check is whether replay caused script execution in the victim browser. \tool performs this check with browser-side instrumentation. It installs a \emph{browser execution sentinel} in the victim page before the page loads and gives the sentinel a fresh single-use token for the current validation attempt. Specifically, we use Chrome DevTools Protocol (CDP) to create a function in the global Javascript context of the victim browser, we create a random secret value (the ``token'') and give this token to the attacker script, run the attacker script, and then check whether the function was invoked with that secret value.
The exploration agent is instructed to create an attacker script and payload that will invoke the sentinel function with the token.
\tool accepts a finding only when the victim browser reports that page script invoked the sentinel with that token.

The resulting verdict has a narrow but important meaning: under the enforced replay, environment, and role constraints, script executed in the victim browser. The sentinel alone does not prove exploitability; the preceding constraints ensure that the observed execution corresponds to the submitted proof of concept and the application's threat model.

\subsection{Acceptance Semantics}
\label{sec:harness:summary}

The mechanisms above define the acceptance semantics of \tool. An accepted proof of concept is not merely evidence that script executed somewhere; it means that, from a clean verification environment, the payload was delivered through the application interface under the attacker role and executed in the victim role's browser during replay, and the attacker had a great deal of control over the payload. This is the attacker-to-victim XSS condition that \tool is designed to validate. The verdict is deterministic with respect to these replay inputs: for a fixed web application snapshot, threat model, and submitted proof of concept, \tool runs the same constrained replay and accepts only if the victim-browser sentinel observes execution.

The combination of constraints prevents agents from ``cheating.''
Environment isolation prevents state or source changes from exploration from affecting replay. PoC constraints ensure that the attacker acts only as a remote HTTP client and that the victim side contains only ordinary browser actions. Role separation binds the planting and rendering steps to the declared threat model. Verdict binding accepts only a fresh browser-side execution signal from the victim browser. Together, these constraints remove the capabilities used by the reward-hacking behaviors in Section~\ref{sec:motivation}, rather than trying to detect them after the fact.
\section{Implementation}
\label{sec:implementation}

We implement \tool in approximately 5K lines of Python as a verification layer that can be paired with an exploration agent or framework. The artifact includes both the implementation and the evaluation benchmark.


\tool uses two application instances per target web application: one for exploration and one for verification. Each instance runs in its own Docker container, and the attacker script runs in a separate sandbox container connected only to the verification instance’s private network. The sandbox can send HTTP requests to the verification instance, but has no shell access to either application instance. \tool reduces startup cost by assigning each validation check a fresh instance from a pre-warmed pool.

Each benchmark target ships configuration scripts rather than a bundled copy of the application source: a download script that fetches the source at a pinned version, and a build script that constructs and starts the application Docker image with baseline accounts. Targets whose threat model spans multiple roles, such as a low-privilege attacker and an administrator victim, include a seed script that creates these additional accounts. A manifest records the attacker and victim roles used by the threat model, based on the application’s security documentation when available. On the verification instance, \tool applies the diverged credential seeding of Section~\ref{sec:harness:roles}, replacing each configured non-attacker password with a verifier-only value so that credentials observed during exploration are stale during verification.

The submission interface represents the two-role proof of concept using three fields: an attacker script, a victim URL, and victim actions. The attacker script defines the attacker side of the replay; the victim URL and victim actions define the victim side. During validation, \tool authenticates the victim browser according to the manifest and sanitizes the submitted victim URL by removing embedded credentials and configured authentication parameters. Each victim action pairs a user-interaction verb, such as click, fill, select, scroll, or wait, with a CSS selector. The verifier sets up the browser execution sentinel through the Chrome DevTools Protocol (CDP)~\cite{chromedevtoolsprotocol}. \tool records whether the sentinel observes execution on page load or only after a victim action, which helps maintainers triage the replayed proof of concept.
\section{Evaluation}
\label{sec:evaluation}

\subsection{Experiment Setup}
\label{sec:eval:setup}

\paragraphtitle{Agents and models}
All experiments use the same base exploration agent: Claude Code through the Agent SDK, with Claude Opus 4.6 as the underlying model. We use Claude Opus 4.6 because it was a stable Opus model for agentic coding and security-testing workflows at the time of our experiments. Each run starts from the target web application repository and a fixed task prompt asking the agent to find XSS vulnerabilities. The agent receives no target-specific hints, such as bug location, CVE identifier, or trigger condition. We fix the agent and model across configurations so result differences reflect the verifier setup rather than model changes.
Each run has a fixed per-target token budget (\perTargetBudget) and a five-hour wall-clock limit. We compute cost from billed token usage using the provider’s published API rates.

\paragraphtitle{Benchmark curation} 
We evaluate \tool on two benchmark suites totaling \benchTotal targets from real-world open-source web applications. The targets span PHP, JavaScript/TypeScript, Python, Java, Go, Ruby, and C\#, and range from \starsRange GitHub stars, with a median of \medStars stars.

The \emph{unknown-vulnerability discovery suite} consists of the latest released versions of \zdTotal applications. Any confirmed XSS in this suite therefore affects released software at the time of evaluation, although it may overlap with undisclosed or unfixed issues already known to maintainers.

The \emph{known-vulnerability recovery suite} consists of \odTotal targets from \odDistinct different applications with already-disclosed and already-patched XSS vulnerabilities. These disclosed CVEs define the ground-truth bugs for recall: a run succeeds only if \tool recovers a working exploit for the corresponding bug.
Existing CVE benchmarks are not directly reusable for this evaluation: CVE-Bench and CVE-Bench v2.0 targets broad web-CVE exploitation with standard attack goals, while our evaluation requires XSS-specific recovery under attacker-to-victim role semantics and the no-leak controls below~\cite{cvebench2025, cvebench2}. 

\paragraphtitle{Preventing Data Contamination} 
To evaluate true discovery rather than memorization, we enforce a strict no-leak protocol. First, to prevent \emph{pretraining leakage}, the known-vulnerability suite includes only CVEs disclosed after the underlying model’s August 2025 training cutoff. Second, to prevent \emph{runtime leakage}, we strip disclosure-derived information from the agent’s environment: we check out the last vulnerable commit before the patch, remove later git history, strip CVE identifiers and advisory text from source and build output, and block external network access.

We construct each target web application from its repository source code and package it as a resettable container. Each package includes build and seed scripts, along with a manifest recording the application URL, credentials, documented attacker and victim roles, trusted-HTML-author features, and, for recall targets, the ground-truth CVE. We derive roles from the repository’s security documentation when available. This packaging lets \tool build the application, seed accounts, replay submitted proofs of concept, and reset the target between verification checks.

\subsection{RQ1: Real-World Vulnerability Discovery}
\label{sec:eval:rq1}

We first evaluate whether \tool can discover previously unknown XSS vulnerabilities in real-world open-source web applications. We compare its effectiveness and precision with the baselines by running all tools on the latest released versions of the \zdTotal applications in the unknown-vulnerability discovery suite. Each run uses the same exploration agent, model, prompt, budget, and target setup described in \S\ref{sec:eval:setup}.

\paragraphtitle{Baselines}
We compare \tool against two classes of baselines. The first class consists of three black-box XSS scanners: ZAP~\cite{ZAP}, YuraScanner~\cite{yurascanner2025}, and Black Widow~\cite{eriksson2021blackwidow}. ZAP and Black Widow are non-LLM black-box scanners, while YuraScanner is an LLM-assisted scanner that uses an LLM to exercise application functionality before applying XSS tests. All three test the deployed application without source-code access. The second baseline is the same Claude Code agent without \tool’s verifier: its prompt asks the agent to find XSS vulnerabilities, generate a proof of concept for each candidate, and verify it against the running application before reporting. Because the baselines do not always produce replayable proofs of concept, we manually reconstructed and replayed reported candidates when needed.

\begin{table}[t]
\centering
\caption{\textbf{Effectiveness on the unknown-vulnerability discovery suite} (\zdTotal targets). \emph{Reachable} counts targets whose attack surface the approach could exercise. Precision is undefined when no reports are produced.}
\label{tab:zeroday}
\small
\setlength{\tabcolsep}{3.4pt}
\begin{tabular}{@{}lrrrr@{}}
\toprule
\textbf{Approach} & \textbf{Reachable} & \textbf{Reports} & \textbf{TP} & \textbf{Precision} \\
\midrule
ZAP                                & 31/\zdTotal & 22 & 0  & 0\% \\
YuraScanner                        & 26/\zdTotal & 10 & 0  & 0\% \\
Black Widow                        & 22/\zdTotal & 0 & 0  & n/a \\
Claude Code (self-judge)           & \zdTotal/\zdTotal & 27 & 10  & 37\% \\
\tool                             & \zdTotal/\zdTotal & 30 & 30  & 100\% \\
\bottomrule
\end{tabular}
\end{table}

\paragraphtitle{Results}
Table~\ref{tab:zeroday} summarizes the unknown-vulnerability discovery results. 
We count a report as a true positive only if it demonstrates a real security risk: attacker-supplied script executes in the victim browser under the intended attacker-to-victim threat model. Otherwise, we count it as a false positive. 
On the \zdTotal targets, \tool reported 30 distinct findings, all true positives; each was manually verified under the intended roles and stored as a replayable proof of concept.

The same Claude Code agent using self-judgment was much less reliable. It reported 27 findings. When we reconstructed and replayed them in a real browser, 10 were true positives and 17 were false positives. Among the 10 true positives, only four were directly replayable as reported; the other six required manual repair to produce an executable PoC. The false positives mostly treated reflection as execution: the agent often fetched a page with \texttt{curl}, observed its payload in returned HTML, and reported success even though no JavaScript executed in a browser.

The black-box scanners found no XSS vulnerabilities.
They had problems exploring the website: for 19 of 50 applications, ZAP was not able to exercise any application web page or endpoint, other than startup/login/health-check pages (YuraScaner: 24/50, Black Widow: 28/50).
In contrast, white-box agentic exploration was able to exercise web pages for all 50 applications. This reachability gap reflects a known limitation of black-box web scanning: scanners normally test applications from the outside, and crawler-driven exploration struggles with complex workflows, authenticated or administrative states, JSON API interactions, and single-page frontends whose inputs are constructed dynamically by client-side JavaScript. Even for applications they could crawl, the scanners produced no true positives. Their only outputs were lower-confidence candidates: ZAP raised 22 potential-XSS alerts, and YuraScanner raised 10 HTML-injection candidates. We manually replayed these candidates with executing payloads in a real browser, but none produced JavaScript execution.

Together, these results show that white-box agentic exploration is effective for reaching and exercising real-world XSS attack surfaces, but verification must be independent of the agent's own judgment. In this evaluation, \tool found XSS vulnerabilities while admitting no false positives.

\begin{table*}[t]
\centering
\caption{\textbf{Representative XSS findings from the unknown-vulnerability discovery suite confirmed and disclosed by \tool.} Each finding was confirmed by browser execution under the declared trust boundary and reported to maintainers. 
}
\label{tab:findings}
\scriptsize
\setlength{\tabcolsep}{3.5pt}
\begin{tabularx}{\textwidth}{@{}l l l X l c l l@{}}
\toprule
\textbf{Application} & \textbf{Stack} & \textbf{Type} & \textbf{Finding / sink} & \textbf{Trust boundary} & \textbf{Sev.} & \textbf{Disclosed} & \textbf{Status} \\
\midrule
cvat               & Python     & Stored       & markdown editor                                                  & User $\rightarrow$ User        & C 9.0       & 2026-04-23 & Acknowledged \\
nicegui            & Python     & Reflected    & \texttt{X-Forwarded-Prefix} via Jinja \texttt{|\,safe}           & Guest $\rightarrow$ User    & C 9.0       & 2026-06-01 & Acknowledged \\
frappe             & Python     & Stored       & report-view link title (\texttt{innerHTML})                      & User $\rightarrow$ Admin   & H 8.7       & 2026-04-23 & Patched \\
changedetection.io & Python     & Stored       & watch-history (\texttt{?html=true} served inline)                & User $\rightarrow$ Admin   & H 8.7       & 2026-04-23 & Acknowledged \\
orchardcore        & C\#        & Stored       & recipe / shortcode import                                        & User $\rightarrow$ Admin   & H 8.1       & 2026-04-23 & Acknowledged \\
cvat               & Python     & Stored       & asset-upload \texttt{Content-Type} spoof, served inline          & User $\rightarrow$ User & H 7.6       & 2026-05-31 & Acknowledged \\
wallos             & PHP        & Stored       & notification settings                                            & Admin $\rightarrow$ Admin  & M 6.4       & 2026-04-23 & Patched \\
Docmost            & Node/TS    & Stored & path-traversal on public image endpoint                          & User $\rightarrow$ Guest   & M 5.4       & 2026-04-23 & Patched \\
filebrowser        & Go         & Stored & uploaded SVG served inline                                       & User $\rightarrow$ Guest &  M 5.4      & 2026-04-22 & Patched \\
navidrome          & Go         & Stored       & \texttt{sanitizeText} bypass in welcome message                  & User $\rightarrow$ Guest     & M 5.4       & 2026-04-22 & Patched \\
changedetection.io & Python     & Stored       & tag-modal name (\texttt{innerHTML})                              & User $\rightarrow$ Admin   & M 5.4       & 2026-04-22 & Acknowledged \\
audiobookshelf     & JavaScript & Stored       & podcast description unsanitized on create                        & Admin $\rightarrow$ User       & M 4.8       & 2026-04-22 & Acknowledged \\
\bottomrule
\end{tabularx}
\end{table*}

\paragraphtitle{Impact triage and disclosure}
Of the 30 vulnerabilities found by \tool, 24 were previously unknown and 6 were known but still unpatched in the latest releases. We disclosed all previously unknown issues to maintainers; at the time of writing, 12 have been acknowledged and patched or scheduled for upstream patching.

Table~\ref{tab:findings} summarizes representative exploitable findings from the unknown-vulnerability discovery suite. We manually triaged each confirmed vulnerability by XSS type, affected roles, and estimated CVSS severity. XSS type distinguishes stored from reflected XSS, while affected roles capture the attacker-to-victim relationship required for exploitation. Overall, the findings are not mere payload reflection or attacker-local execution: they are exploitable XSS vulnerabilities in which attacker-controlled input reaches and executes in a distinct victim’s browser.

\begin{figure}[t]
\begin{lstlisting}[style=prettybox,language=JavaScript]
@Get('attachments/img/:attachmentType/:fileName')
async getLogoOrAvatar (@Param('fileName') fileName, ...){
    // Validation uses only the trailing path segment
    const base = path.basename(fileName, path.extname(fileName));
    if (!isValidUUID(base)) 
        throw new BadRequestException();

    // File lookup uses the full requested path.
    const filePath = ${getAttachmentFolderPath(attachmentType, workspace.id)}/${fileName};
    return res.send(await this.storageService.read(filePath));
}
\end{lstlisting}
\caption{\textbf{Vulnerable Docmost image handler} (CVE-redacted, simplified). The snippet shows how validation-to-lookup mismatch enables path traversal via the public image endpoint.}
\label{fig:docmost}
\end{figure}

\begin{figure}[t]
\begin{lstlisting}[style=prettybox,language=pocsh]
# ATTACKER SCRIPT
curl $TARGET_URL/api/auth/login -d '{...}'
PID=$(curl $TARGET_URL/api/pages/create -d '{...}')
# upload dummy; reuse its id as the SVG file name
ID=$(curl $TARGET_URL/api/files/upload
-F file=@dummy.txt -F pageId=$PID)
# upload SVG as "$ID.svg"; response gives its storage dir
DIR=$(curl $TARGET_URL/api/files/upload
-F "file=@xss.svg;filename=$ID.svg"
-F pageId=$PID)

# VICTIM URL
/api/attachments/img/avatar/..%2Ffiles%2F$DIR%2F$ID.svg
# VICTIM ACTIONS
[]
\end{lstlisting}
\caption{\textbf{\tool proof of concept for the Docmost finding} (simplified). The attacker uploads the SVG under a UUID-shaped name (\texttt{\$ID.svg}) and notes its storage directory (\texttt{\$DIR}); the victim URL traverses from the avatar folder into that directory so the trailing segment passes the UUID check; the action list is empty because the SVG executes on render.}
\label{fig:poc}
\end{figure}

\paragraphtitle{Illustrative example: a Docmost SVG XSS (CVE-redacted)}
\tool discovered a previously unknown XSS vulnerability in Docmost, an open-source collaborative wiki and documentation platform. The issue arises from an unsafe composition of two features: Docmost allows SVG files as ordinary page attachments, while its avatar/logo endpoint is intended to inline-serve only safe raster images such as JPG or PNG. As shown in Figure~\ref{fig:docmost}, this endpoint checked only that the requested file’s base name had UUID format and did not reject parent-directory components earlier in the path. An attacker could therefore upload an SVG attachment containing script and retrieve it through the avatar/logo endpoint using a URL-encoded traversal path whose final segment still passed the UUID check. This caused the endpoint to serve attacker-controlled SVG through an inline image-serving path where SVGs were not intended to appear. When a victim opened the crafted URL as a document, the browser executed the SVG script. The maintainers acknowledged the issue and patched the affected code.

This finding illustrates both the value and the verification risk of white-box agentic discovery. The agent could inspect the endpoint, connect the attachment-upload workflow to the public avatar/logo serving path, and infer the exploit constraint needed to route an uploaded SVG through an endpoint intended for safe raster images. A black-box scanner would have to discover this multi-step composition through interaction alone. However, the security claim is not simply that an SVG can execute, but that an attacker can upload the SVG through the application, have it served through the unintended public inline path, and trigger execution in a separate victim browser. An unconstrained white-box agent could still make the browser execute script through shortcuts, such as direct storage writes, filesystem changes, browser injection, or privileged credentials. These shortcuts show execution, but not the harmful attacker-to-victim XSS path.

\tool preserves the white-box discovery advantage while requiring final evidence to follow that attacker-to-victim path. The submitted proof of concept is shown in Figure~\ref{fig:poc}. From a clean verification snapshot, the attacker side interacts with Docmost only over HTTP: 
it logs in, creates content, uploads a benign file to obtain a storage identifier, uploads the malicious SVG attachmentch, and constructs the public URL that routes it through the avatar/logo endpoint. 
The victim side runs in a separate browser context with no attacker session or exploration state; it only opens the crafted public URL and performs constrained user actions.
The confirmed PoC therefore establishes the required XSS chain: the attacker supplies the payload through the application, the public endpoint serves it in an executable context, and a distinct victim browser executes it.

\result{RQ1}{On latest releases, \tool reported 30 confirmed XSS findings with no false positives; 24 are previously unknown vulnerabilities, and 12 have already been acknowledged by maintainers and patched or scheduled for upstream patching. Without \tool's independent verifier, the same agent reported 27 successes, but only 10 are real XSS vulnerabilities.}

\begin{table}[t]
\centering
\caption{\textbf{Detection and recovery on the known-vulnerability recovery suite} (\odTotal targets). Detection counts any confirmed XSS exploit; recovery requires a confirmed exploit matching the labeled CVE.}
\label{tab:recall}
\small
\begin{tabularx}{\columnwidth}{@{}Xr@{}}
\toprule
\textbf{Outcome} & \textbf{Targets} \\
\midrule
\textbf{Confirmed (TP)}, detection 49\%, recovery 36\%                & \textbf{22 / \odTotal} \\
\quad recovered the target's labeled CVE                 & 10 \\
\quad recovered it, plus other bug(s)                    & 6 \\
\quad found a different real bug, not the labeled CVE    & 6 \\
\midrule
\textbf{Not found (FN)}                              & 23 / \odTotal \\
\midrule
\textbf{False positives (FP)}                            & 0 / \odTotal \\
\bottomrule
\end{tabularx}
\end{table}

\subsection{RQ2: Known-Vulnerability Recovery}
\label{sec:eval:rq2}

We next evaluate \tool on applications with disclosed XSS CVEs, using the CVEs only as ground truth rather than as hints. For each target, we run \tool on the pre-patch version under the same no-leak controls: disclosure text, post-fix history, and external vulnerability sources are unavailable to the agent. This setup measures discovery on vulnerable code without allowing the public disclosure to guide exploration.

We report two target-level outcomes. \emph{Detection} means that \tool confirms at least one real XSS exploit in the frozen application. \emph{Recovery} is stricter: at least one confirmed exploit must match the labeled CVE ground truth. Thus, a target can be detected but not recovered if \tool finds a different real XSS vulnerability in the same pre-patch code.

\paragraphtitle{Results}
Table~\ref{tab:recall} summarizes the results. \tool achieved 49\% detection and 36\% recovery on known-vulnerability targets. Every accepted finding was a real, reproducible XSS exploit; there were no false positives. Confirmed exploits either recovered the labeled CVE or exposed a different real XSS vulnerability in the same frozen codebase; we separately reviewed these additional findings and reported previously unreported ones to maintainers.

Targets without a confirmed exploit are false negatives of the end-to-end discovery pipeline. These misses came from exploration-side limitations: either the agent did not identify the vulnerable code path without disclosure-derived hints, or identified a plausible issue but failed to construct a PoC accepted by \tool. \tool intentionally focuses on high-confidence vulnerability reports, so that we don't waste developers' time with false positives.
Future improvements to exploration could potentially reduce false negatives without weakening the no-false-positive acceptance criterion.

\result{RQ2}{On the known-vulnerability recovery suite, \tool rediscovered the labeled CVE in 36\% of targets and detected real XSS exploits in 49\%. Every accepted finding was a reproducible browser-confirmed exploit: \tool had no false positives.}

\subsection{RQ3: Reward-Hacking Prevention}
\label{sec:eval:rq3}

We evaluate the effectiveness of verification (at preventing reward hacking and false positives) and of end-to-end vulnerability detection (at finding vulnerabilities), under different types of verification mechanisms.
We conduct a cumulative ablation on a subset of 27 applications from the unknown-vulnerability suite (Table~\ref{tab:ladder}). All configurations use the same exploration agent and targets; only the verifier setup changes.

\begin{table}[t]
\centering
\caption{\textbf{Cumulative ablation of \tool mechanisms} (27 targets). Starting from an agent that runs and checks its own work, each row adds one \tool\ mechanism.}
\label{tab:ladder}
\small
\begin{tabularx}{\columnwidth}{@{}Xrrr@{}}
\toprule
\textbf{Configuration} & \textbf{Reports} & \textbf{TP} & \textbf{Precision} \\
\midrule
Agent runs the check, agent judges              & 11 & 5  & 45\%\\
\;\;$+$ Browser-bound verdict                   & 111 & 10 & 9\% \\
\;\;$+$ Environment isolation                   & 67  & 13  & 19\% \\
\;\;$+$ PoC constraints                         & 32  & 14  & 44\% \\
\;\;$+$ Role separation (full \tool)            & 14  & 14  & 100\% \\
\bottomrule
\end{tabularx}
\end{table}

\emph{Agent self-judgment.}
The baseline lets the agent judge its own success. It produces low precision (45\%) because the agent often treats weak evidence, such as reflected payload text or self-observed page changes, as script execution. Additionally, out of the 5 TPs, only two were directly replayable as reported; the others  required manual repair.
It finds few vulnerabilities because it often fools itself into (wrongly) thinking it has found a vulnerability and terminates prematurely, before finding a real vulnerability.

\emph{Browser-bound verdict.}
Replacing self-judgment with a browser-bound verdict avoids conflating reflection with execution, but we observed that reward hacking was very frequent: the agent often directly runs server-side commands to modify the database or filesystem to trigger the browser sentinel, without exploiting any XSS vulnerability.
Precision dropped to 9\%, showing that browser execution alone is insufficient when a white-box agent can manipulate the environment.

\emph{Environment isolation.}
Isolation resets the application before each verification attempt, removing state planted through privileged or out-of-band channels during exploration. This reduces contaminated reports and improves precision to 19\%.

\emph{PoC constraints.}
PoC constraints allow arbitrary code only on the attacker side as a remote HTTP client and restrict victim behavior to typed browser actions. This prevents verifier self-injection and raises precision to 44\%.

\emph{Role separation.}
Role separation requires the exploit to cross the declared attacker-to-victim trust boundary. This removes findings that rely on the agent acting as a trusted or more privileged user; with the full \tool configuration, all 14 accepted reports are true positives, yielding 100\% precision.

The ablation shows that no single mechanism is sufficient. Self-judgment mistakes weak evidence for execution, and a browser-bound verdict alone may be manipulated through verifier inputs. Environment isolation, PoC constraints, and role separation respectively remove state contamination, verifier self-injection, and wrong-user execution, converting a noisy browser check into a high-precision verifier.

\result{RQ3}{The ablation shows that \tool's verifier defenses progressively remove reward-hacking false confirmations: precision changes from 45\% with agent self-judgment to 9\%, 19\%, 44\%, and finally 100\% as browser-bound verdicts, environment isolation, PoC constraints, and role separation are added. The initial drop shows that browser execution alone is insufficient; the full configuration accepts 14 findings, all true positives.}
\section{Discussion}
\label{sec:discussion}

\paragraphtitle{Execution evidence reduces review cost} \tool makes it easier for developers to review vulnerability findings because it shifts the focus of human review from exploit confirmation to impact assessment. With prior tools, reviewers must first determine whether there is a real vulnerability. In contrast, \tool only reports a vulnerability if there is a proof that a payload can execute in the victim's browser, and makes it easy to reproduce this. Reviewers therefore do not need to reconfirm basic execution.
This reduced the review task from checking whether the vulnerability is exploitable to deciding which findings were worth disclosing.

\paragraphtitle{Software vulnerabilities beyond XSS}
Similar methods could be used to verify other vulnerability classes, when success has an observable execution signal.
For instance, SQL injection could be validated through timing behavior or out-of-band database callbacks, server-side request forgery could be validated through a controlled listener, and path traversal could be validated through a planted file containing an unguessable token.
The verifier must be specialized for each class. It must define the attacker-controlled input, the victim or server action, and the success signal.
Our methods to counter reward hacking would be applicable to all of these vulnerability classes.
Our current prototype assumes a headless Chromium browser with DevTools Protocol support, which leaves mobile WebViews and targets with debugging disabled outside our current scope.

\paragraphtitle{Threats to Validity}
\label{sec:eval:tov}
Our verifier could accept false vulnerability claims if the role assignment is incorrect (most consequentially, an ``attacker'' who is in fact a trusted HTML author), or if the evaluated snapshot contains state not present in a clean deployment.
To mitigate this threat, we derive roles from each application's security documentation, exclude roles that are documented as trusted to inject HTML, and freeze role assignments before each run;
whether external defenses (e.g., a web application firewall) would block exploitation remains a deployment-specific manual judgment.
We only evaluated one configuration, Claude Code with Claude Opus 4.6.
Other models or harnesses might perform differently, but our verifier would remain applicable.
\section{Related Work}
\label{sec:related-work}

Prior XSS techniques differ in discovery and verification.
We compare them using the requirements in Section~\ref{sec:motivation}: execution-grounded validation (R1), tamper-resistant validation (R2), and attacker-exploitable validation (R3).

\textbf{Static, source-guided, and exploit-generation analysis} inspects code or program structure without relying only on end-to-end black-box exploration.
Classic taint analyzers such as Pixy~\cite{jovanovic2006pixy}, QL/CodeQL-style query frameworks~\cite{avgustinov2016ql}, XSS-specific systems such as Splendor~\cite{su2023splendor}, local path-persistent fuzzing systems such as XSSky~\cite{shi2025xssky}, and exploit-generation systems such as NAVEX~\cite{alhuzali2018navex} and Ardilla~\cite{kiezun2009ardilla}, together with static detection of second-order web vulnerabilities~\cite{dahse2014secondorder}, expose candidate source-to-sink flows and can reason across application logic.
However, such flows may be infeasible, unreachable, or only exploitable under roles unavailable to a realistic attacker.
Static findings therefore remain candidate claims that require downstream validation, and false positives and triage cost remain central challenges~\cite{bessey2010,johnson2013,owaspbenchmark}. They do not by themselves satisfy R1 or R3.

\textbf{Black-box dynamic} scanners validate vulnerabilities by interacting with the application through its web interface. Tools such as ZAP~\cite{ZAP} and Burp~\cite{burp} inject payloads and look for reflection or execution, while research systems improve the crawler, scanner realism~\cite{drakonakis2023rescan}, or the XSS verifier.
Black Widow~\cite{eriksson2021blackwidow} performs black-box data-driven web scanning with emphasis on XSS, Black Ostrich~\cite{blackostrich2023} uses string solvers to handle input-validation constraints, Dancer in the Dark~\cite{kirchner2024dancer} targets blind XSS, YuraScanner~\cite{yurascanner2025} uses LLM-based task execution with Black Widow's XSS detection engine, and Spider-Scents~\cite{olsson2024spiderscents} adds grey-box database awareness for stored XSS.
Earlier work established detection and exploit validation for DOM-based and persistent client-side XSS~\cite{lekies2013flows,melicher2018domsday,steffens2019locals,parameshwaran2015dexterjs} and evolutionary black-box XSS fuzzing~\cite{duchene2014kameleonfuzz}, and showed that sanitization can be bypassed through DOM mutation (mXSS)~\cite{heiderich2013mxss}, motivating verifiers grounded in real execution rather than reflection alone.
Recent work drives DOM-XSS discovery deeper through webpage-interaction fuzzing and URL-component synthesis~\cite{sabino2026swipe}.
These systems can satisfy R1 when validation is grounded in browser execution, but black-box reach is limited by authentication, role-specific state, multi-step workflows, input constraints, and stored-then-rendered paths~\cite{doupe2010johnny,doupe2012enemy,guo2025evocrawl}. They also often leave attacker-to-victim exploitability and replayable PoC construction outside the acceptance condition.

\textbf{LLM-assisted and agentic security systems} use program-analysis structure, tool use, or interactive environments to improve vulnerability discovery.
LLM-assisted systems such as IRIS~\cite{ICLR2025_582d4e27}, LLMxCPG~\cite{llmxcpg2025}, and GPTScan~\cite{sun2024gptscan} (for smart-contract logic flaws) use program-analysis structure to identify candidate vulnerabilities.
Tool-using and agentic systems such as PentestGPT~\cite{pentestgpt2024}, EnIGMA~\cite{enigma2025}, HPTSA~\cite{zhu-etal-2026-teams}, and Revelio~\cite{hou2026revelio} show the promise of agents for security tasks, while SWE-agent~\cite{sweagent2024} demonstrates the broader utility of agent-computer interfaces for software-engineering tasks.
Industry systems such as XBOW~\cite{xbow2025} provide further evidence of practical interest in agentic security workflows.
EvoCrawl~\cite{guo2025evocrawl} and Atropos~\cite{guler2024atropos} improve web-application exploration and testing infrastructure, and CVE-Bench~\cite{cvebench2025} complements these systems by evaluating agents on real-world web-application CVEs.
Offensive-security agent benchmarks such as Cybench~\cite{zhang2025cybench}, NYU CTF Bench~\cite{shao2024nyuctf}, and CyberGym~\cite{wang2026cybergym} further track agent exploitation capability.
However, once such an agent is granted white-box access, a verification problem arises: the agent can influence application state, browser profile, validation scripts, callback channels, database contents, or the submitted PoC, satisfying R1 while violating R2 or R3.
This mirrors broader evidence of reward hacking, benchmark-validity failures, and tool-using agents being subverted through untrusted inputs~\cite{baker2025monitoring,zhu2025abc,debenedetti2024agentdojo}.
\section{Conclusion}
\label{sec:conclusion}

This paper presents \tool, a verification framework for white-box agentic XSS discovery. 
\tool accepts a finding only when the submitted proof of concept reproduces script execution in a victim browser under a realistic threat model, so each confirmed finding provides evidence of a real exploit rather than a reward-hacking artifact produced by an underconstrained verifier. 
\tool reported 30 confirmed XSS findings on latest releases, including 24 previously unknown vulnerabilities. On known-vulnerability recovery targets, it recovered the labeled CVE in 36\% of targets and confirmed real exploits in 49\%. We responsibly disclosed all previously unknown vulnerabilities found by \tool.  

\section*{Acknowledgment}
This research was supported by the Noyce Foundation and gifts from Accenture, Amazon, AMD, Anyscale, Broadcom, Google, IBM, Intel, Intesa Sanpaolo, Lambda, Lightspeed, Mibura, NVIDIA, Samsung SDS, SAP, by the U.S. Department of Energy, and the Defense Advanced Research Projects Agency (DARPA) under Agreement No. HR00112590134. Hao Wang is grateful for the support from Amazon AI Fellowship. Any opinions, findings, and conclusions expressed in this material are those of the authors and
do not necessarily reflect the views of the sponsors. We would also like to thank Yu-Lin Uriah Tsai, Yibo Peng, Aadi Shah, Xinyi Cynthia Wang, and Xuzhe Alina Zhi for their insightful feedback.

\bibliographystyle{IEEEtran}
\bibliography{ref}

\end{document}